\documentclass[prb,twocolumn,superscriptaddress,showpacs]{revtex4-1}
\usepackage{graphicx}
\usepackage{epsfig}
\usepackage{amssymb,amsmath,amsfonts,hyperref}
\usepackage{wasysym}
\usepackage{latexsym}
\usepackage{color}

\newcommand{\nn}{\nonumber}

\newcommand{\be}{\begin{eqnarray}}
\newcommand{\ee}{\end{eqnarray}}

\begin{document}
\title{Coulomb phase diagnostics as function of temperature, interaction range and disorder}
\author{Arnab Sen}
\affiliation{\small{Max-Planck-Institut f\"{u}r Physik komplexer Systeme, 01187 Dresden, Germany}}
\author{R. Moessner}
\affiliation{\small{Max-Planck-Institut f\"{u}r Physik komplexer Systeme, 01187 Dresden, Germany}}
\author{S. L. Sondhi}
\affiliation{\small{Department of Physics, Princeton University, Princeton, New Jersey 08544, USA}}

\begin{abstract}
The emergent gauge field characteristic of the Coulomb phase of spin ice betrays its existence via pinch points in the spin structure factor ${\cal{S}}$ in reciprocal space which takes the form of a transverse projector ${\cal{P}}$ at low temperature: ${\cal{S}} (q) \sim {\cal{P}} \sim q^2_\bot/q^2$. We develop a theory which establishes the fate of the pinch points at low and high temperature, for hard and soft spins, for short- and long-ranged (dipolar) interactions, as well as in the presence of disorder. We find that their detailed shape can be used to read off the relative sizes of entropic and magnetic Coulomb interactions of monopoles in spin ice, and we resolve the question why pinch points have been experimentally observed for Ho$_{1.7}$Y$_{0.3}$Ti$_2$O$_7$ even at high temperature in the presence of strong disorder.
\end{abstract}

\date\today
\vskip2pc

\maketitle

{\it{Motivation}}: The study of `topological' states of matter has been one of the central enterprises of condensed matter physics over the years, encompassing as they do fractional quantum Hall states, spin liquids or topological insulators. They are distinct from conventional phases exhibiting local order -- e.g. a ferromagnet or charge-density wave -- but also from simple disordered phases.

The central experimental challenge is to diagnose their existence in more than
a negative way, such as the {\em absence} of order. Especially for states with
mobile charged excitations, non-local probes such as
conductivity~\cite{shotnoise} or interferometry~\cite{Camino} have been used
to detect edge states or fractionalised excitations; in their absence,
diagnostics have often resorted to general thermodynamic arguments, like the
presence of power laws in thermodynamic quantities~\cite{balents}, 
which while highly suggestive have no direct link to the nature of the effective degrees of freedom.

\begin{figure}
{\includegraphics[width=\hsize]{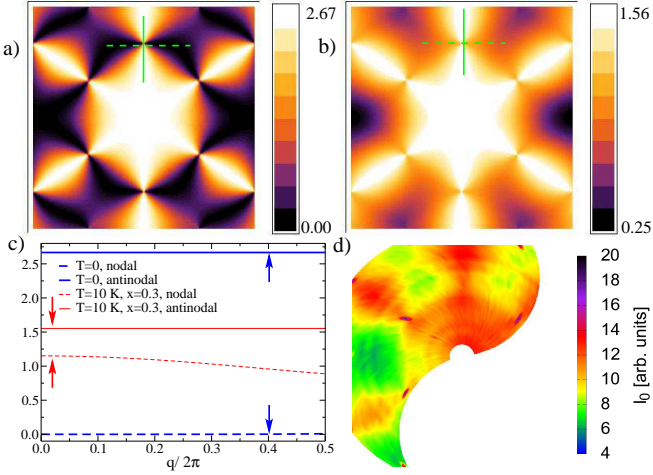}} 
\caption{(a) $\tilde{\mathcal{S}}(\mathbf{q})$ (see Eq~\ref{def}) calculated in the [hhk] plane for the ice states at $T=0$ (b) $\tilde{\mathcal{S}}(\mathbf{q})$ calculated for
  Ho$_{2-x}$Y$_x$Ti$_2$O$_7$ at $T=10$ K  and
  $x=0.3$ (c) Diffuse scattering along nodal direction (dotted line in (a) and (b)) and
  perpendicular to it (antinodal direction, bold line in (a) and (b)) for both the cases near the pinch point $(0,0,4\pi)$. $q$ is measured relative to the pinch point. Note the weak pinch point discontinuity (indicated by arrows) even at high $T$ and disorder. (d) Experimentally observed diffuse scattering from Ref~\onlinecite{LJChang} for
  Ho$_{2-x}$Y$_x$Ti$_2$O$_7$ at $T=10$ K  and
  $x=0.3$. In the three diffuse scattering plots, $q_x \in [-4\pi,4\pi]$ and $q_z \in [-6\pi, 6\pi]$ and measured in units of $a_{\rm cubic}^{-1}$.}
\label{fig1}
\end{figure}

Spin ice, a three-dimensional frustrated magnetic material on the pyrochlore
lattice made of corner-sharing tetrahedra, exhibits a topological
Coulomb phase~\cite{CMS}. A pinch-point motif in the spin structure factor (Fig
\ref{fig1}(a)) has been identified as a diagnostic of the emergent gauge field
characteristic of that phase, and analysed in considerable
detail~\cite{Bramwell,Fennell,Morris,kadowaki,Isakov,Henley,Shannon}. It
emerges as a consequence of a local constraint, the ``ice rule'', imposed by
the interactions, that each tetrahedron have vanishing total (pseudo-)spin
$S_{\XBox}=\sum_{i \in \XBox}S_i= 0$~\cite{Isakov, Henley, Youngblood}.
Excited tetrahedra violating this constraint carry an emergent gauge
charge~\cite{Isakov, Henley, RMSLS, Hermele, Henleyreview} and interact
through both an entropic and a magnetic (energetic)~\cite{Henleyreview,claudio_nature} Coulomb interaction.

It turns out that the pinch points persist to regimes far away from the
Coulomb phase, most strikingly to high temperatures \cite{isakov_icerules} 
even in the presence of strong dilution of the magnetic Holmium ions by
non-magnetic Yttrium 
ions~\cite{LJChang}. This calls for a theory determining under which
conditions pinch points provide what information.

We hence devise a theory for the structure factor across a range of settings
(temperature, nature of spin, range of interaction, level of disorder) and
find that they encode much useful and interesting information. A combined
analysis of the correlations along the two axes of the pinch point which we
call nodal~\cite{Henley} and antinodal (see Fig~\ref{fig1}) reveals this
information, in particular the discontinuity where they cross at the pinch
point itself, as well as the nature of the vanishing of the nodal correlators,
the importance of which has been  emphasized by Henley~\cite{Henleyreview} and
Bramwell~\cite{BramwellST}. We find the possibility of qualitatively distinct
correlation functions for gauge charges and spin correlators: the latter are
enhanced, while the former are screened, as long-ranged interactions are
introduced. The qualitative enhancement is weakened, but does not disappear,
at high temperature, even in the presence of disorder; while at low $T$, one
can read off the relative size of entropic and magnetic charges. 

Our analysis thus contributes not only to the study of diagnostics of topological phases, as well as their interplay of disorder, but also has connections to the unusual correlations induced by dipolar interactions, which are also of much current interest in the study of cold atoms~\cite{Buchler}.

The core of our analysis consists of a formulation of the large-$n$ expression for correlations of the soft-spin model~\cite{Isakov} in terms of quantities natural to the emergent gauge degrees of freedom, which then straightforwardly generalises to the other settings. This may be of more general methodological interest. Here it provides an attractively simple yet complete picture in terms of (unusual forms of) Debye screening.

The analysis starts from the Hamiltonian \cite{Bramwell}
\be
H &=& \sum_{(ij)} H_{ij} S_i S_j = \frac{J}{3} \sum_{(ij)} S_i V_{ij} S_j \nn \\
&+& D a^3 \sum_{(ij)}\left(\frac{\hat{e}_i \cdot
    \hat{e}_j}{|r_{ij}|^3}-\frac{3(\hat{e}_i \cdot \mathbf{r}_{ij})(\hat{e}_j
    \cdot \mathbf{r}_{ij})}{|r_{ij}|^5} \right) S_i S_j
\label{eq1}
\ee 
where $(ij)$ denotes the sum over pairs of spins, $\frac{J}{3}$ and $D$ parametrise the strengths of nearest-neighbour
exchange and dipolar interactions, respectively and $V_{ij}=\Gamma_{ij}+\delta_{ij}$, where $\Gamma$ is the adjacency matrix of the pyrochlore
lattice. We will also have occasion to consider an `ideal' dipolar interaction, which differs from Eq.~\ref{eq1} only by terms vanishing as $\mathcal{O}(r_{ij}^{-5})$ with distance, the matrix form of which corresponds to the projector $\cal P$ onto the ground states of the ice model~\cite{isakov_icerules}. 

The pyrochlore lattice is made of up and down tetrahedra labelled by $\eta = \pm 1$, each up tetrahedron meeting down tetrahedra at its four vertices and vice-versa. It can be viewed as a fcc Bravais lattice with an up tetrahedron (four-site basis) at each fcc site. The unit cell is then a conventional cubic cell of side $a_{\rm cubic}=2\sqrt{2}a$, where $a$ is the nearest neighbour distance. $S_i$ are (hard or soft) spin variables pointing along local easy axes $\hat{e}_\alpha$ in the $\alpha = 1 \dots 4$ [111] directions pointing from the centers of the up tetrahedra to the vertices. The centers of all tetrahedra form a bipartite diamond lattice with bond length $a_d=\sqrt{3}a_{\rm cubic}/4$. 

{\it{Nearest-neighbour soft spins}:} The large-$n$ approach~\cite{Canals} works demonstrably well for the nearest-neighbor (NN) Heisenberg ($n=3$) model at low $T$~\cite{Isakov,Henley,Conlon}, and for the Ising ($n=1$) model at $T=0$~\cite{Isakov}.  The solution requires the
eigenvalues and eigenvectors of $V_{ij}$. In momentum space,
\begin{displaymath}
V(\mathbf{q}) =
\left( \begin{array}{cccc}
2 & 2c_{yz} & 2c_{xz} & 2c_{xy}   \\
2c_{yz} & 2 & 2c_{\overline{xy}} & 2c_{\overline{xz}}  \\
2c_{xz} & 2c_{\overline{xy}} & 2 & 2c_{\overline{yz}}  \\
2c_{xy} & 2c_{\overline{xz}} & 2c_{\overline{yz}} & 2 
\end{array} \right)
\end{displaymath}
where $c_{ab} = \cos \left(\frac{q_a+q_b}{4} \right)$ and $c_{\overline{ab}} =
\cos \left(\frac{q_a-q_b}{4} \right)$. Its spectrum consists of four bands $v_{\mu}(\mathbf{q}) =0,0,4 \mp 2 \sqrt{1+\Upsilon}$ with
$\Upsilon=c^2_{xy}+c^2_{\overline{xy}}+c^2_{yz}+c^2_{\overline{yz}}+c^2_{xz}+c^2_{\overline{xz}}-3$. 

There are thus two flat bands at zero energy  and two bands (one gapless, quadratic acoustic and one optical) with energies $\tilde{v}_{3,4}(\mathbf{q})= (J/3)v_{3,4}(\mathbf{q})$ for the NN model ($D = 0$). The eigenvectors in the flat bands satisfy the ice rules, i.e. the sum of spins on each tetrahedron vanishes.

The spin correlations are expressed in the diagonal basis of
$V(\mathbf{q})$ using a unitary transformation $U(\mathbf{q})$. 
Then, $\langle S_{\alpha} (\mathbf{q}) S_{\beta}(-\mathbf{q}) \rangle$ equals:
\be
\sum_{\mu=1}^2
\frac{U_{\alpha \mu}(\mathbf{q})U^{\dag}_{\mu \beta}(\mathbf{q})}{\lambda(T)} 
&+&\sum_{\mu=3}^4 U_{\alpha \mu}(\mathbf{q})U^{\dag}_{\mu \beta}(\mathbf{q}) \left(
  \frac{T}{\tilde{v}_{\mu}(\mathbf{q})+\lambda(T) T}\right) \nn \\
=\sum_{\mu=1}^2
\frac{U_{\alpha \mu}U^{\dag}_{\mu \beta}}{\lambda(T)} 
&+&\sum_{\mu=3}^4 U_{\alpha \mu}U^{\dag}_{\mu \beta} \left(
  \frac{\mathcal{Q}^2}{v_{\mu}(\mathbf{q})+\frac{a^2_{\rm cubic}\kappa^2}{8}}\right)
\label{eq2}
\ee
where the $\mathbf{q}$ dependence of $U$ is suppressed for brevity in the
second line in Eq~\ref{eq2}. The length constraint $\langle S_i^2 \rangle =1/n$ per component is imposed by the Lagrange multiplier $\lambda (T)$. 

The re-writing in the second line in Eq~\ref{eq2}, ($\mathcal{Q}^2
=T/(J/3)$, $a^2_{\rm
  cubic}\kappa^2=8\lambda(T) T/(J/3)$),  will help us interpret results at finite $T$ in terms of (gauge) charges $S_{\XBox} \neq 0$. $\kappa^{-1}$ equals the charge-charge correlation (screening) length. This can be seen by calculating $\langle S_{\XBox}(0) S_{\XBox}(r) \rangle$. For $r \gg a$, this equals:
\be
\langle S_{\XBox}(0) S_{\XBox}(r) \rangle \sim -\eta_0 \eta_r\frac{\mathcal{Q}^2 a^2_{\rm cubic}\kappa^2}{4\pi} \frac{\exp(-r\kappa)}{r/a_{\rm cubic}}~.
\label{eq3}
\ee 
Note that the charge-charge correlation
function is isotropic in space unlike the spin-spin~\cite{Isakov} correlation
function. Also, when the screening length is large, i.e. $a_{\rm cubic}\kappa \rightarrow 0$, the charge density $\langle S^2_{\XBox} \rangle$ approaches the ``bare'' value $\mathcal{Q}^2$ and $\lambda \rightarrow \frac{n}{2}(1+0.22411 \mathcal{Q}^2 n)$.  

From Eq~\ref{eq2}, we can also calculate the structure factor
$\mathcal{S}(\mathbf{q}) =\langle \sum_{\alpha=1}^4 S_{\alpha}(\mathbf{q})
\sum_{\beta=1}^4 S_{\beta}(-\mathbf{q}) \rangle $. Its pinch points at $T=0$ are due to the singularities in the
$U(\mathbf{q})$ matrix~\cite{Canals}.
 Concentrating on small $(q_x,q_x,q_z)$ near the pinch
point at $(0,0,4\pi)$ for clarity, $\mathcal{S}(\mathbf{q})$ can be written as: 
\be
 \frac{4q_x^2}{q_x^2+q_z^2} \left(
  \frac{1}{\lambda(T)} \right)
+\frac{4q_z^2}{q_x^2+q_z^2} \left( \frac{ \mathcal{Q}^2}{v_3(\mathbf{q})+\frac{a_{\rm cubic}^2\kappa^2}{8}} \right)
\label{eq4}
\ee
where we omit the contribution of the gapped optical branch, which is fully analytic. 

At $T=0$, only the flat bands contribute, leading to pinch points. Moreover, when $q_x=0$, the structure factor is precisely zero near the pinch point. At finite $T$, there is non-zero diffuse scattering in this nodal direction~\cite{Henley}, proportional to $\mathcal{Q}^2$, the width of which is controlled by $a_{\rm cubic}\kappa$, the inverse screening length. In the perpendicular antinodal direction, the scattering arises from the flat band states and is present even at $T=0$ (Fig~\ref{fig1}(c)). The discontinuity in
$\mathcal{S}(\mathbf{q})$ is washed out at finite $T$ and the structure factor is
analytic at the pinch point. In real space, spin correlations now decay exponentially
with a correlation length $\xi/a_{\rm cubic} =
\frac{1}{2}\sqrt{(J/3)/nT}$ at low $T$. Thus, the spin correlations and the
charge correlation decay with the same correlation length in the NN case.
 
The charge correlations at low $T$ can be understood as Debye-H\"uckel (DH) screening~\cite{DH} of charges $S_{\XBox}$. Unusually, their size is not ``quantised'' (because of the continuous nature of the soft spins) and their Coulomb interaction is {\em entropic}. The charges effectively live on the dual diamond lattice sites which are the centers of the tetrahedra of the pyrochlore lattice. The varying number of spin configurations consistent with a given charge configuration leads to the entropic Coulomb interaction $\mathcal{K}a_d \sum_{i,j}Q_i Q_j/r_{ij}$ where $\mathcal{K}=nT/(2\sqrt{3}\pi)$ and $Q_i = \eta_i (S_{\XBox})_i$. The screening length equals $a_{\rm cubic}^2 \xi^{-2}_{DH}=(8 \sqrt{3}\pi \mathcal{K}\rho \langle Q^2 \rangle)/T$ within standard DH theory. For soft spins, each tetrahedron hosts a charge of mean size $\mathcal{Q}$ at low $T$: indeed, setting $\rho \langle {Q}^2 \rangle=T/(J/3)$ gives $\xi_{DH}=\kappa^{-1}$.    

{\it{Nearest-neighbour (hard) Ising spins}:}
Ising spins cannot simply be described in terms of modes with eigenvalues $\tilde{v}_{\mu}(\mathbf{q})$ because the unit length constraint couples the modes non-linearly. However, the rewritten form, Eq. \ref{eq2}, does generalise straightforwardly in the sense of an effective theory. The low-$T$ properties of spin ice can be described in terms of a dilute set of mobile, Coulombically interacting charges on the dual diamond lattice. The crucial difference to soft spins is that now these charges are quantised, and hence gapped. The parameters $\mathcal{Q}^2 $ and $a_{\rm cubic}\kappa$ are fixed by the properties of these interacting charges, while the Lagrange multiplier $\lambda(T)$ enforces the spin length constraint on average. Note that $v_{\mu}(\mathbf{q})$ are the eigenvalues of the Laplacian operator $\Gamma_{ij}-z \delta_{ij}$ on the
pyrochlore lattice, where $z$ is the coordination number, (with a further shift of $8\delta_{ij}$ to make the eigenvalues positive-definite) which turns into $\nabla^2$ in continuum. 

We first consider the NN Ising model. At low $T$, the (exponential) majority of defects have $S_{\XBox} = \pm 2$ so that $S_{\XBox} = \pm 4$ may safely be ignored. The fraction of defective tetrahedra $\rho$ is itself exponentially suppressed and equals $2\exp(-2J/3T)$.  Since $\langle S^2_{\XBox}\rangle =4\rho$, this fixes $\mathcal{Q}^2 = 4\rho$. The Lagrange multiplier $\lambda(T=0)=1/2$ since $n=1$ for Ising spins. The weight in the flat bands decreases according to $\mathcal{Q}^2$ at finite $T$, which we take into account by $\lambda(T)=1/2+\alpha(T)\rho$. Specifying $\mathcal{Q}^2$ and $\lambda(T)$ fixes the inverse screening length $a^2_{\rm cubic}\kappa_e^2=16 \rho+32 \alpha(T) \rho^2$ for the NN Ising model as we require the pinch points to vanish at any $T \neq 0$. $\alpha(T)$ can be determined numerically from the unit-length constraint and equals $0.44822$ as $T \rightarrow 0$. The leading term in $a^2_{\rm cubic}\kappa_e^2$ exactly reproduces the entropic screening length calculated in Ref~\onlinecite{CC} using DH theory where the gapped charges ($Q=\pm 2$) interact with an entropic Coulomb interaction of strength $\mathcal{K}=T/2\sqrt{3}\pi$.

\begin{figure}
{\includegraphics[width=\hsize]{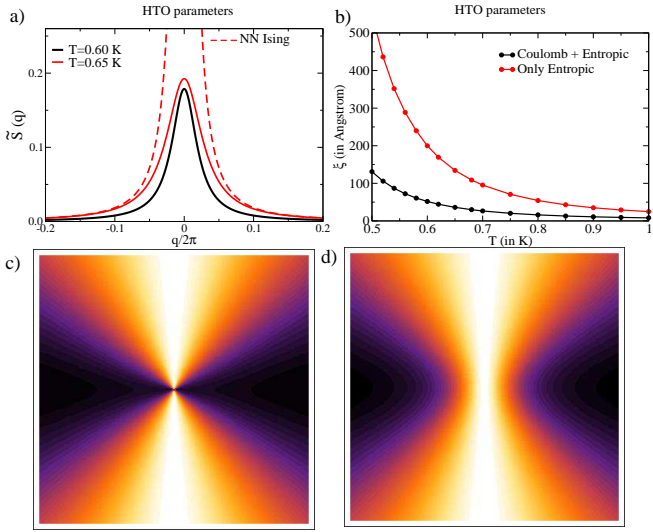}}
\caption{(a) $\tilde{\mathcal{S}}(\mathbf{q})$ calculated along the nodal line near the pinch point $(0,0,4\pi)$ for both the NN Ising model and dipolar spin ice (using HTO parameters). $q$ is measured relative to the pinch point. The dotted line shows the scattering for the NN Ising model using the value of $\rho$ obtained at $T=0.65$ K with HTO parameters. The dotted line meets the antinodal line ($\approx 2.66$) at $q=0$.(b) ``Correlation length'' $\xi$ defined using the HWHM of the diffuse
  scattering calculated  using HTO parameters (black points) in the nodal direction. Red points show
  the corresponding NN Ising result using the $\rho$ obtained with HTO parameters. (c) $\tilde{\mathcal{S}}(\mathbf{q})$ near the pinch point $(0,0,4\pi)$ in the [hhk]
  plane for HTO parameters at $T=0.65$ K. Note the discontinuity even at finite $T$. 
(d) $\tilde{\mathcal{S}}(\mathbf{q})$ calculated using the same
  $\rho$ in the NN Ising case. The discontinuity is rounded off. Farther away from the pinch points, (c) and (d) agree. In both the scattering plots, $q_x/2\pi \in [-0.05, +0.05]$ and $q_z/2\pi \in [1.95, 2.05]$. }
\label{fig2}
\end{figure}

The pinch points now acquire a finite width in $\mathbf{q}$ space, determined by $a_{\rm cubic}\kappa_e$, of order $|q| \sim \sqrt{\rho}$ for Ising spins~\cite{Henleyreview,BramwellST} (see Figs~\ref{fig2}(a),(d)) instead of $\sim \sqrt{T/J}$ for soft spins. This width can be observed in the half-width at half maximum (HWHM) of the diffuse scattering along the pinch point's nodal direction. In spin ice, the Ising spins point along the easy-axes and thus, the experimentally relevant structure factor $\tilde{\mathcal{S}}(\mathbf{q})$ has additional phase factors,~\cite{Fennell} and equals 
\be
\sum_{\alpha,\beta}\langle S_{\alpha}(\mathbf{q})S_{\beta}(-\mathbf{q}) \rangle \left( \frac{\hat{e}_{\alpha} \cdot (\hat{e}_{||} \times \mathbf{q})}{|\hat{e}_{||} \times \mathbf{q}|} \frac{\hat{e}_{\beta} \cdot (\hat{e}_{||} \times \mathbf{q})}{|\hat{e}_{||} \times \mathbf{q}|}\right) 
\label{def}
\ee
where $\hat{e}_{||}=\frac{1}{\sqrt{2}}(1,-1,0)$ and $\mathbf{q}$ lies in the [hhk] plane. We display $\tilde{\mathcal{S}}(\mathbf{q})$ in the discussions of diffuse scattering rather than $\mathcal{S}(\mathbf{q})$. Note that in three dimensions, the DH screening length $\rho^{-1/2}$ scales differently from the average distance between the thermal defects, $\sim \rho^{-1/3}$. As in the soft spin case, the NN Ising case also has both the spin-spin correlations and the charge-charge correlations decaying with the same correlation length $\sim \rho^{-1/2}$         at finite (low) $T$.

{\it{Dipolar spin ice}:} 
We now turn to dipolar spin ice where long-ranged interactions between the
spins, $D \neq 0$, play an important role. Due to the dipolar interactions,
the charges $S_{\XBox} \neq 0$ now have an {\it additional} magnetic Coulomb
interaction.~\cite{claudio_nature} This can be conveniently seen by using the
dumbbell model~\cite{claudio_nature} where each spin (point dipole) is
replaced by an extended dipole of size $a_d$ placed on each bond of the dual
diamond lattice. Then, up to correction terms that decay as
$\mathcal{O}(1/r_{ij}^5)$ for large $r_{ij}$, the interactions in Eq~\ref{eq1}
can be re-written as  
\be
H_d= \frac{2\sqrt{2}}{3\sqrt{3}}D a_d \sum_{i>j}\frac{Q_i Q_j}{r_{ij}}+ \Delta \sum_i (Q_i/2)^2
\label{eq5}
\ee 
where $\Delta = \frac{2J}{3}+\frac{8}{3}\left(1+\sqrt{\frac{2}{3}} \right)D$.

Thus, the charges interact with both an entropic and a magnetic Coulomb interaction and the total  coupling is $\mathcal{K} = T/(2\sqrt{3}\pi) + 2\sqrt{2}D/(3\sqrt{3})$. The magnetic Coulombic DH screening length equals $a^2_{\rm cubic}\kappa_{M}^2 = \frac{64 \sqrt{2}\pi}{3T} \rho D $. Here, $\rho$ needs to be calculated self-consistently~\cite{CC} and equals $2\exp(-\Delta/T)$ as $T \rightarrow 0$. Thus, the fraction of defective tetrahedra now depends on both $J$ and $D$.

The parameters of the generalized equation are now 
$ \mathcal{Q}^2  =4\rho$, $\lambda(T)=1/2+\alpha(T)\rho$ and $\kappa^2=\kappa_e^2+\kappa_{M}^2$ where $\alpha(T)$ is determined from the unit length constraint. The parameters go smoothly to the NN Ising values when $J \gg D$. The charge-charge correlators calculated from this generalized equation have the correct DH form when both interactions are present.

Screening is now enhanced as the magnetic Coulomb interaction does not vanish with $T$. The contribution of the acoustic band is then suppressed by $\mathcal{O}(T/D)$ for wave-vectors $|q| \ll \kappa_{M}$ near the pinch point as $T \rightarrow 0$, because $\rho/\kappa^2_{M} \sim T/D$. Thus, the pinch points survive but are weakened. More specifically, the discontinuity at the pinch points, when approaching from the nodal and the antinodal directions respectively, scales as $1-(1+4\sqrt{2}\pi D/3T)^{-1}$ at low $T$. Secondly, the HWHM for the scattering in the nodal direction at the pinch point scales as $\sqrt{(1+4\sqrt{2}\pi D/3T)}\sqrt{\rho}$. Both the discontinuity at the pinch point and the HWHM in the nodal direction show that the gauge charges interact both energetically and entropically: $(4\sqrt{2}\pi D)/(3T)$, the ratio of the magnetic and the entropic Coulomb couplings, can be probed directly in these quantities.

Fig~\ref{fig2}(a) shows the calculated diffuse scattering along the nodal
direction at low $T$ using Ho$_2$Ti$_2$O$_7$ (HTO)
parameters~\cite{bramewellexp} of $D=1.41$ K and $J=-1.56$
K. Fig~\ref{fig2}(c), shows the survival of the pinch point at finite $T$
using HTO parameters at $T=0.65$ K. A corresponding ``correlation length'',
associated with the charge-charge correlation, can be extracted from the HWHM in the nodal direction (Fig~\ref{fig2}(b)). In the same figure, we show the correlation length if the calculated defect density $\rho$ is used as the input parameter
in the NN Ising case, where the charges only interact entropically. The growth of the correlation length is slower in the dipolar spin ice case because of the magnetic Coulomb screening.   

Thus, the picture that emerges is that the charge-charge correlations are again exponentially decaying in real space with a scale $a^2_{\rm cubic}\xi ^{-2} \sim \left(1+\frac{4\sqrt{2}\pi D}{3T} \right)\rho$ signaling the sparse defects of the Coulomb phase. Scattering along the nodal directions near the pinch points, driven entirely by these charges, reflects the screening of these charges which is now enhanced by the presence of the magnetic Coulomb interactions. However, the spin-spin correlations have a power-law tail which is dipolar and decays as $1/r^3$ {\em at all temperatures}, signaling the presence of long-range interactions and not of a Coulomb phase. The last feature can already be seen from the soft-spin problem where the spins interact through the `ideal' dipolar interactions~\cite{isakov_icerules} $H_{ij} = \mathcal{P}_{ij} = \delta_{ij}- \langle S_i S_j \rangle_{T=0,n=1}$, where $\langle S_i S_j \rangle_{T=0,n=1}$ is calculated using Eq~\ref{eq2} at $T=0$ and $n=1$. Then the spin-spin correlations $\langle S_{\alpha}(\mathbf{q})S_{\beta}(-\mathbf{q}) \rangle$ become
$\left( \frac{n}{\lambda(T)}-1 \right)  \langle S_{\alpha}(\mathbf{q})S_{\beta}(-\mathbf{q}) \rangle_{T=0,n=1}+\left(\frac{2}{n}-\frac{1}{\lambda(T)}\right)\delta_{\alpha \beta}$, with $\lambda(T)=\frac{nT-1+\sqrt{1+n^2T^2}}{2T}$. The charge-charge correlations are however extremely short-ranged and non-zero only for two tetrahedra sharing a common site.

This observation holds the key to explaining the presence of well-defined pinch points at high temperature $T = 10$ K and strong chemical disorder in experiments on spin ice~\cite{LJChang} diluted with non-magnetic Y, Ho$_{2-x}$Y$_x$Ti$_2$O$_7$, where the constraint $S_{\XBox}=0$ must be massively violated. These pinch points are entirely due to the form of the long-range nature of the dipolar interactions on the pyrochlore lattice, which mimic the spin-spin correlations at $T=0$.~\cite{isakov_icerules} This can be seen explicitly in the leading term of a high-temperature series expansion generalized to include disorder where the probability of a site being occupied by a magnetic  Ho$^{3+}$ ion is $(1-x/2)$:
\be
\langle S(r_1) S(r_2) \rangle&=& \left(1-\frac{x}{2} \right)^2 \frac{\sqrt{2}\pi D}{3T} \langle S(r_1) S(r_2) \rangle_{T=0,n=1} \nn \\
&-&\left(1-\frac{x}{2} \right)^2 \left(\frac{J}{3}+\left( \frac{5-\frac{\sqrt{2}\pi}{3}}{3}\right) D\right) \frac{\Gamma_{r_1,r_2}}{T} \nn \\
\langle S(r)^2 \rangle &=& \left(1-x/2 \right) ~~.
\label{eq6} 
\ee
From this, we calculate $\tilde{\mathcal{S}}(\mathbf{q})$ for Ho$_{2-x}$Y$_x$Ti$_2$O$_7$ at
$x=0.3$ at a high temperature of $T=10$ K (Fig~\ref{fig1}(b)). The diffuse scattering pattern is similar to the experimental one (Fig~\ref{fig1}(d)).
Note that the discontinuity at the pinch point is weak in $D/T$, and the ``nodal'' correlations are broad and large (Fig~\ref{fig1}(c)). However, higher-order terms in $1/T$ will not necessarily preserve the dominance of the pinch point features and instead lead to rich low-temperature physics including a dilution dependent ``residual entropy''~\cite{KeSchifferMoessner} and the possibility of glassiness.~\cite{Gaulin, review}

{\it{Conclusions}}:
Defects with gauge charge ($S_{\XBox} \neq 0$) in both NN and dipolar models can be generated by thermal excitations or chemical disorder. For $D \neq 0$, these carry both entropic and magnetic Coulomb charges, while they carry only entropic charge in the NN model. This leads to a qualitative change in the spin correlations which can be observed in equilibrium experiments  probing the pinch points. We have provided a simple and explicit computational and conceptual framework encompassing all these cases. 

{\it{Acknowledgements}}:
We thank C.~Castelnovo (also for helpful comments on the manuscript),
K.~Damle, M.~Gingras, K.~Shtengel and the authors of Ref~\onlinecite{LJChang}
for useful discussions, and the latter for Fig~\ref{fig1}(d).
This research
was supported in part by the National Science Foundation under Grant No. DMR
10-06608 (SLS).

\end{document}